\documentclass[twocolumn]{article}

\usepackage{blindtext} % Package to generate dummy text throughout this template

\usepackage[sc]{mathpazo} % Use the Palatino font
\usepackage[T1]{fontenc} % Use 8-bit encoding that has 256 glyphs
\usepackage{microtype} % Slightly tweak font spacing for aesthetics

\usepackage[english]{babel} % Language hyphenation and typographical rules

\usepackage[hmarginratio=1:1,top=16mm,columnsep=13pt,left=18mm]{geometry} % Document margins
\usepackage[hang, small,labelfont=bf,up,textfont=it,up]{caption} % Custom captions under/above floats in tables or figures
\usepackage{booktabs} % Horizontal rules in tables

\usepackage{lettrine} % The lettrine is the first enlarged letter at the beginning of the text

\usepackage{enumitem} % Customized lists
\setlist[itemize]{noitemsep} % Make itemize lists more compact

\usepackage{abstract} % Allows abstract customization
\usepackage{titlesec} % Allows customization of titles
\renewcommand\thesection{\Roman{section}} % Roman numerals for the sections
\renewcommand\thesubsection{\roman{subsection}} % roman numerals for subsections
\titleformat{\section}[block]{\large\scshape\centering}{\thesection.}{1em}{} % Change the look of the section titles
\titleformat{\subsection}[block]{\large}{\thesubsection.}{1em}{} % Change the look of the section titles

\usepackage{fancyhdr} % Headers and footers
\usepackage{titling} % Customizing the title section

\usepackage{hyperref} % For hyperlinks in the PDF

\usepackage[]{algorithm2e}
\usepackage{multirow}
\usepackage{graphicx}
\usepackage{subcaption}
\usepackage{xspace}
\usepackage{xcolor}

\usepackage{listings}
\usepackage{amssymb}
\usepackage{fancyvrb}

\newcommand{\patchtovec}{\textsf{commit2vec}\xspace}
\newcommand{\ptov}{\patchtovec}
\newcommand{\wordtovec}{\textsf{word2vec}\xspace}
\newcommand{\codetovec}{\textsf{code2vec}\xspace}

\pretitle{\begin{center}\LARGE\bfseries} % Article title formatting
\posttitle{\end{center}} % Article title closing formatting
\title{\textsc{Commit2Vec}: Learning Distributed Representations of Code Changes} % Article title
\author{%
\textsc{Roc\'io Cabrera Lozoya, Arnaud Baumann, Antonino Sabetta, Michele Bezzi}\\[1ex]
\normalsize SAP Security Research % Your institution
}
\date{} % Leave empty to omit a date
\renewcommand{\maketitlehookd}{%
\begin{abstract}
\noindent Deep learning methods, which have found successful applications in fields like
 image classification and natural language processing, have recently been
 applied to source code analysis too, due to the enormous amount of freely available source code (e.g., from open-source software repositories).

In this work, we elaborate upon a state-of-the-art approach to the representation of source code that uses information about its syntactic structure, and we adapt it to represent source code \textit{changes} (i.e., commits). We use this representation to classify security-relevant commits.

Because our method uses transfer learning (that is, we train a network on a ``pretext task'' for which abundant labeled data is available, and then we
use such network for the target task of commit classification, for which fewer labeled instances are available), we studied the impact of pre-training the network using two different pretext tasks versus a randomly initialized model.

Our results indicate that  representations that leverage the structural information obtained through code syntax outperform token-based representations.
Furthermore, the performance metrics obtained when pre-training on a loosely related pretext task with a very large dataset ($>10^6$ samples) were surpassed when pretraining on a smaller dataset ($>10^4$ samples) but for a pretext task
that is more closely related to the target task.
\end{abstract}

}

\begin{document}

\hspace*{-1.1\columnwidth}
\enlargethispage{3\baselineskip}
\begin{minipage}{\textwidth}
\begin{center}
{\LARGE
\textsc{Commit2Vec}: Learning Distributed Representations of Code Changes}\\[4mm]
{\large\bf [PRE-PRINT]} \\[6mm]
Roc\'io Cabrera Lozoya, Arnaud Baumann, Antonino Sabetta, Michele Bezzi
\end{center}

\thispagestyle{empty}
\vspace{5mm}
\small
\begin{abstract}
\noindent Deep learning methods, which have found successful applications in fields like
 image classification and natural language processing, have recently been
 applied to source code analysis too, due to the enormous amount of freely available source code (e.g., from open-source software repositories).

%  In existing approaches, source code changes were treated as documents in
%  natural language processing, potentially ignoring the structured nature of
%  source code.

In this work, we elaborate upon a state-of-the-art approach to the representation of source code that uses information about its syntactic structure, and we adapt it to represent source code \textit{changes} (i.e., commits). We use this representation to classify security-relevant commits.

% We show how our approach, which incorporates syntactic properties of code, results in better classification performance when compared to token-based approaches.

Because our method uses transfer learning (that is, we train a network on a ``pretext task'' for which abundant labeled data is available, and then we
use such network for the target task of commit classification, for which fewer labeled instances are available), we studied the impact of pre-training the network using two different pretext tasks versus a randomly initialized model.

Our results indicate that  representations that leverage the structural information obtained through code syntax outperform token-based representations.
Furthermore, the performance metrics obtained when pre-training on a loosely related pretext task with a very large dataset ($>10^6$ samples) were surpassed when pretraining on a smaller dataset ($>10^4$ samples) but for a pretext task
that is more closely related to the target task.
% These results highlight the importance of adequately choosing the pretext task used for pretraining.

\end{abstract}

\vspace{4mm}
\normalsize
\hrule
\vspace{2mm}
\begin{center}
{\large Citing this paper}
\vspace{2mm}
\end{center}

% This is a pre-print of the paper that appears in the proceedings of the 
% 16th IEEE/ACM International Conference on Mining Software Repositories (MSR), 2019.

\noindent
Please note that this preprint is an early version of an improved and substantially extended
paper that appeared in Springer Nature Computer Science in 2021, and that is freely available
from the website of the publisher (\url{https://doi.org/10.1007/s42979-021-00566-z}).

Please refer to that paper instead of this pre-print; you can cite is as follows: 
\vspace{2mm}
{\tt
\begin{Verbatim}
@Article{CabreraLozoya2021,
    author={Cabrera Lozoya, Roc{\'i}o
        and Baumann, Arnaud
        and Sabetta, Antonino
        and Bezzi, Michele},
    title={Commit2Vec: Learning Distributed Representations of Code Changes},
    journal={SN Computer Science},
    year={2021},
    month={Mar},
    day={19},
    volume={2},
    number={3},
    pages={150},
    issn={2661-8907},
    doi={10.1007/s42979-021-00566-z},
    url={https://doi.org/10.1007/s42979-021-00566-z}
}

\end{Verbatim}
}

\vspace{2mm}
\hrule
\medskip
This work was partly supported by EU-funded
projects \textsc{AssureMOSS} (Grant no.~952647) and \textsc{Sparta} (Grant no.~830892).

\vspace*{\stretch{1}}
\hfill
\includegraphics[width=45mm]{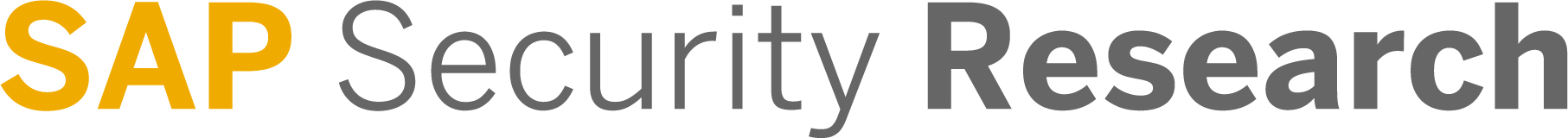}
\raisebox{-6mm}{\includegraphics[width=23mm]{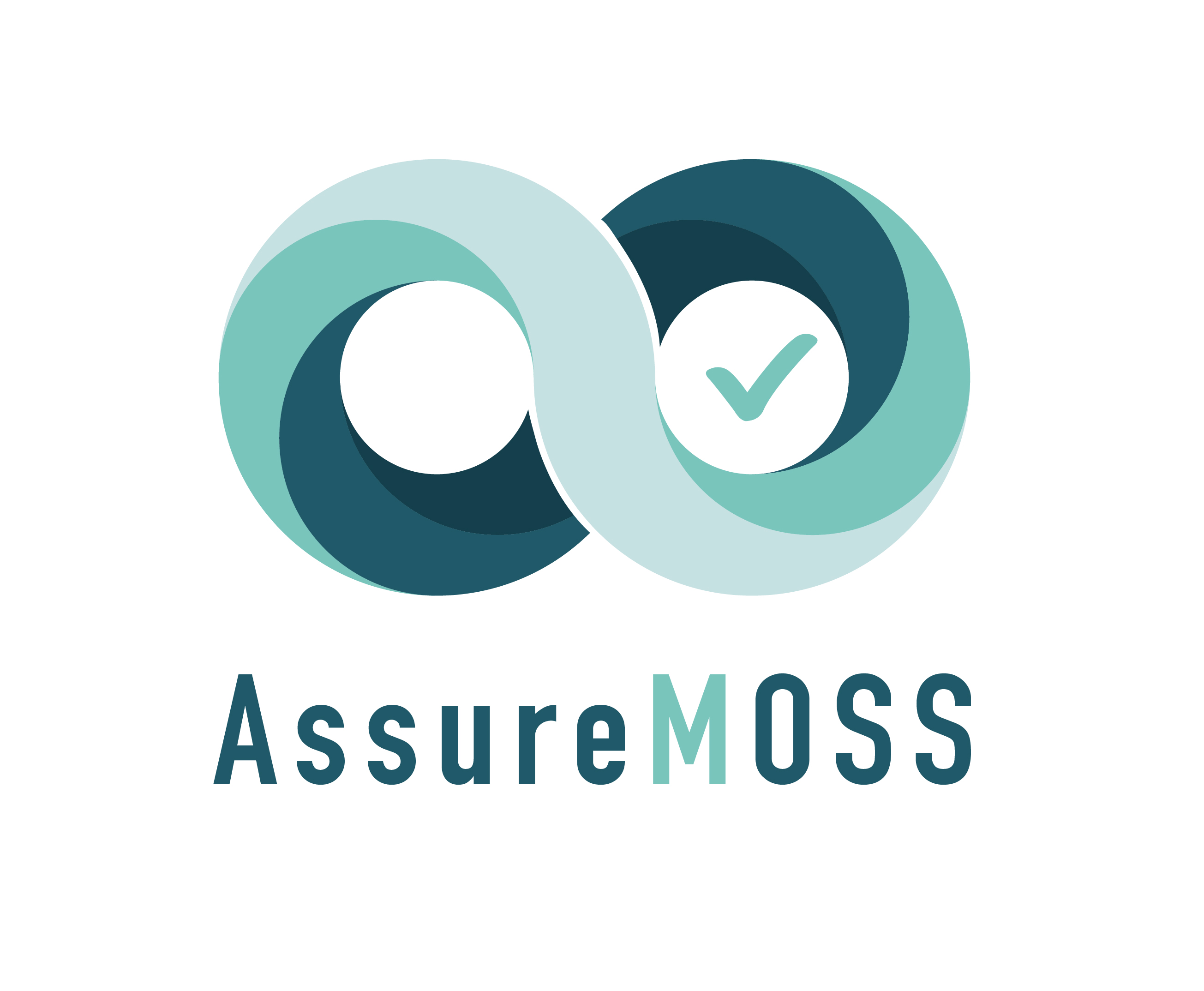}}
\raisebox{-8mm}{\includegraphics[width=23mm]{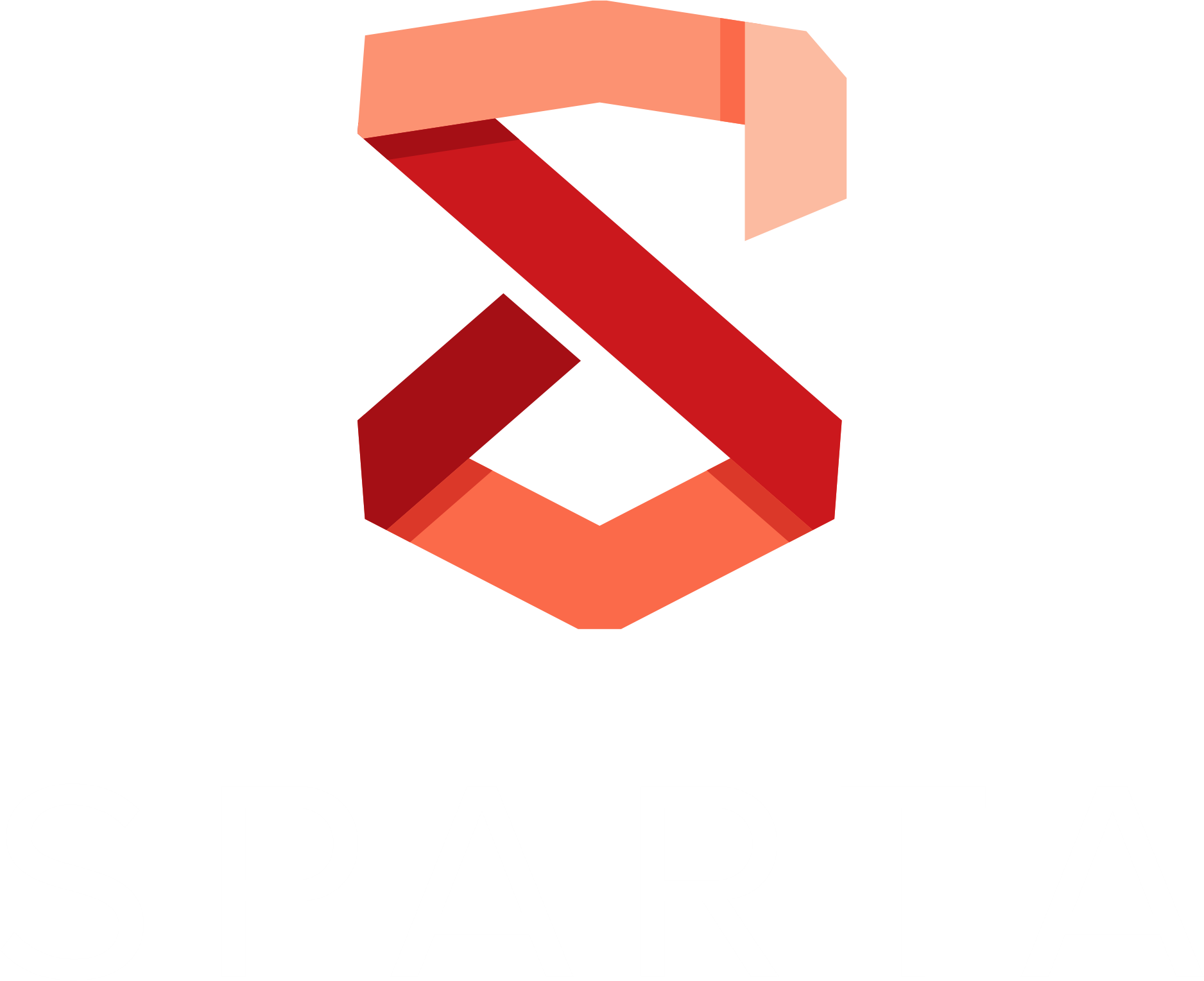}}

% \raisebox{-6mm}{\includegraphics[width=23mm]{logo_assuremoss.png}}\hfill
% \includegraphics[width=4cm]{logo_sap_sr.jpg}\hfill
% \raisebox{-10mm}{\includegraphics[width=23mm]{logo_sparta.png}}

\end{minipage}

% Print the title
\maketitle

\section{Introduction}
\label{sec:intro}

% \as{this is work in progress\ldots}

%  The vulnerability management process of a software with open source components
%  is challenging due its dependence on non-reliable standard sources of
%  advisories and vulnerability data. Previous efforts aimed to reduce this
%  dependency by directly analyzing source code for the automatic detection of
%  commits that are security-relevant. In~\cite{sabetta2018practical}, source
%  code changes were treated as documents in natural language processing,
%  potentially ignoring the structured nature of source code. In this work, we
%  compare the performances of token-based approaches to those incorporating
%  syntactic properties of code into the analysis. We leverage on the work of
%  ~\cite{code2vec} which generates generates distributed representations of code
%  snippets by analyzing and aggregating paths extracted from the abstract syntax
%  tree of the code. We generalize this concept to the representation of
%  \textit{code commits} through code differences and analyze the impact of
%  pre-training a neural network on high volumes of data with different
%  relevances to our target task. \rcl{Our results show that semantic-based
%  representations perform better than token-based representations by XXX and
%  that pre-training on a highly relevant, yet smaller, database outperforms the
%  use of a huge, weakly-related task.}

%--------------------------------------------------

Deep learning methods have been proven successful in a variety of problems,
such as image classification, natural language processing, speech recognition,
and others. More recently there is a growing interest in using similar approaches to
programming-language related tasks~\cite{code2seq,code2vec,surveyCodeEmbeddings,representedits}, using code as the main input source.

To this end, a key prerequisite is the ability to represent the code (or code fragments) as a numerical vector ({\it embedding}), similarly to the \wordtovec~\cite{word2vec} approach for natural
language processing (NLP). Such vectorial representation should have the property of mapping similar instances of code elements onto close points in the
embedding vector space. Using NLP methods to build representations of software code
is meaningful, indeed, as empirically shown in~\cite{hindle2012naturalness}. Source code is
characterized by similar statistical properties as natural language ({\it
naturalness hypothesis}~\cite{surveyAllamanis}, which is not surprising
considering that code is written and read by \emph{humans}, in
addition to being executable by machines). On the other hand, there are
significant differences, since code is written in a programming language, which
is a formal language: it presents minimal ambiguity, large re-use of identical ``sentences'', and reduced robustness to small changes compared to natural language. In addition, the
semantic units of text, as sentences or paragraphs, are typically relatively
short, present a high level of locality, and they rarely used more than one
time in the text. On the contrary, (sequences of) code statements or functions are clearly delimited,
they may be used multiple times in different contexts, and present long range
correlations (i.e., the semantic of a statement can be influenced by other
statements in a somewhat distant part of the code).

For these reasons, beyond NLP-inspired methods, a number of representations
that use the structural nature of code have been proposed, such as using data
flow graphs, control flow graph and abstract syntax trees, and used to perform
tasks as variable and method naming~\cite{Allamanis:2015:SAM:2786805.2786849,code2vec}, clone detection~\cite{clone_detection}, code
completion ~\cite{Raychev:2014:CCS:2666356.2594321,pmlr-v32-santos14}, summarization ~\cite{DBLP:journals/corr/AllamanisPS16}, and algorithm classification~\cite{Mou:2016:CNN:3015812.3016002}.

Due the complexity of the code structure, training a deep learning algorithm to
solve a code-related task needs a large amount of labeled data, which in many
practical cases is not available. Ideally, we would like to build a
low-dimensional representation using a task where a large amount of
labeled data is available (or can be obtained in a relatively inexpensive way), and use the learned representation to solve a different (target) task, where fewer data point are available, but enough to fine tune the model ({\it tranfer learning}).

\noindent The contribution of this paper is twofold.

First, we propose a new model for representing code changes, called \patchtovec,
which, along the lines of~\cite{code2vec}, uses paths from the AST to build a
\emph{representation of code changes}.

Then, we evaluate
the model on an industry-relevant use-case (security commit classification),
where a limited number of samples is available. To this end, we construct datasets to support the pre-training of  embedding models and we refine
the learning by continuing the training on data related to the target task (transfer learning). We show how the performance of a
model can be optmised by learning the code representation on a {\it related}
task (prediction of the priority of issue-tracking tickets) where a large dataset can be easily created,
and fine tune the model to solve the (target) task of identifying security-relevant commits.

%relevance of source code analysis: encouraged by the progress made in outher
%fields, such as image and video processing as well as NLP, researchers have
%started to investigate ways to apply deep learning to the analysis of source
%code.
%
%Impact: create automated tools that improve quality and in particular security.
%
%In this paper we present a novel approach to expressing source code changes
%(e.g., those corresponding to a commit in a source-code repository) that uses a
%distributed representation, suitable to be given in input to machine learning
%models.
%
%In this work we introduce a method to represent source \textit{code commits} and
%use it in a machine learning frameworks to detect security-relevant
%vulnerability fixes.  We represent not only isolated static versions of methods,
%but full commits, which can be composed of changes in multiples methods and in
%multiple files and which implies a dynamic version due to its temporal
%component. We propose an efficient and fully automated solution for the
%representation of \textit{code changes}.

\section{Motivating Scenario}
\label{sec:motivation}
% \as{EXPECTED CONTENTS: security application, oss}

The motivation for this work stems from our interest in the problem of ensuring that the open-source components included in the products of our company are \emph{free from (known) vulnerabilities}. This is among the most urgent security challenges that the software industry is facing~\cite{owasp-top-10}, and has hit the headlines multiple times in the recent past~\cite{equifax,heartbleed}.

The adoption of open-source software (OSS) components by commercial software
developers has increased considerably over the past few years: according
to~\cite{SnykIO}, 80 to 90\% of commercial software includes OSS components.
While this practice allows commercial software vendors to speed up innovation
and decrease development costs, it raises important concerns, especially related
to security. The number of open-source components, the variety of practices
according to which they are developed, and the complexity of their interdependencies
translate into a composition that is difficult to understand and to maintain.

The current approach to vulnerability management consists in keeping track of publicly disclosed vulnerabilities in the National Vulnerability Database (NVD),
which is known to have consistency and coverage issues, among others.
% Whenever a fix for a vulnerability is provided by the OSS
% project, and depending on the severity of the vulnerability, a cost-benefit
% analysis can be done to decide if it is worth updating to the newly released
% version despite the development efforts this would imply.
More in general, this approach does not scale well as the amount of open-source
code developed by the community (and incorporated in commercial products) increases at a steady pace.

To make the problem worse, a study by \textsf{Snyk.io}~\cite{SnykIO} reported that as many as 25\% of the OSS projects fix vulnerabilities without ever disclosing them in an official advisory (so-called \textit{silent fixes}), making it even harder for projects that depend on OSS components to take
informed decisions as to whether they should upgrade to a more recent (non-vulnerable) version.

The urgent need of rigorous practices to ensure the security of the software
supply chain is reflected in the landscape of commercial offerings that have
appeared over the past three years, some of which claim to have extensive
vulnerability data, obtained via monitoring of advisories and project
code repositories.
% \as{mention that new commercial offerings have emerged in the past couple of years providing precisely data services that claim to complement NVD data with
% other vulnerabilities}

% Therefore, there is a community motivation to use machine learning tools to automate the detection of commits that are security-relevant.

%Previous works~\cite{sabetta2018practical,zhou2017automated} \as{[cite others in
%the publication "fight"]; EDIT: actually, not sure we should cite them again...}
%
%The work in~\cite{sabetta2018practical} targets the same task we are tackling:
%the automated classification of commits that are \emph{security-relevant} (i.e.,
%that are likely to fix a vulnerability). They train two independent classifiers
%on the patch introduced by a commit (\emph{Patch Classifier}) and the log
%messages (\emph{Message Classifier}), without relying on information from
%vulnerability advisories. Inspired by the \emph{naturalness}
%hypothesis~\cite{surveyAllamanis,hindle2012naturalness}, both classifiers treat
%their input as documents written in natural language (\emph{code-as-text}), and
%classify them using established Natural Language Processing (NLP) methods. Their
%results are combined with a simple voting mechanism flagging a commit as
%security-relevant if either model does.
%
%Nevertheless, unlike the work in~\cite{sabetta2018practical}, we rely solely on
%the code as a source of information (not using the commit message) to make our
%method robust to \textit{silent fixes}.

Given the amount of open-source code produced and consumed nowadays, it is clear that any approach that is based on manual effort is doomed to fail, and
the role of automated tools (in particular, those based on machine-learning technology) is destined to grow steadily.

The problem of representing code (and code changes) is receing more and more
attention both in academia and industry. However, how to represent source code in a way that can be used effectively in machine learning applications is
an open research question.  Given the huge amount of freely available code
and the recent breakthroughs in other fields such as computer vision, speech
recognition and natural language processing, code analysis is a strong candidate
to be the next application to benefit from deep learning techniques.

% Indeed, software repositories such as GitHub or Gitlab host an enormous amount
% of source code (with an historical record of changes) that is readily available
% for analysis and automated learning. Between September 2016 and end of 2017,
% there have been more than one billion public commits pushed on GitHub. The
% nature of these commits can be vary, ranging from small edits, improvements, the
% introduction of new features or a more critical fix of a bug or a security
% vulnerability.

\section{Background and Related Work}
\label{sec:related}

%Previous works~\cite{sabetta2018practical,zhou2017automated} \as{[cite others in
%	the publication "fight"]; EDIT: actually, not sure we should cite them again...}

In this section we first present some background information on \codetovec~\cite{code2vec}, which we extend and adapt in this work. Then, we briefly
overview related literature.

\subsection{Code2Vec}

The work in \codetovec~\cite{code2vec} uses a bag of aggregated contexts to create distributed representations for single methods, and is trained to predict the name of a method given its body. We adapt the same idea to represent \emph{commits} (code changes), as opposed to individual code snapshots; also, our method represents a commit as a whole, aggregating data of all the methods changed in the commit, whereas \codetovec focuses on individual methods.
In~\cite{code2vec}, to process a method, the model first builds the abstract syntax tree associated to the method, and then creates a set of triplets (named \textit{contexts}) composed of two terminals in the tree and their connecting path. Each path and terminal has an associated embedding of size 128. The three embeddings of the triplet are then concatenated to a single context vector of size 384. Each context vector is propagated to a fully connected layer producing a vector of size 128. Finally, an attention mechanism combines the resulting vectors and creates the \textit{code-vector} of the method. A final layer connected to this code-vector is added in order to compute the loss. In the original paper, a lookup table for the method name and a softmax layer form the last layer. As explained in the next section, we use \codetovec as a building block in our method to represent code changes.

\subsection{Other Works on Code Representation and Classification}

The work in~\cite{sabetta2018practical} is motivated by the same use-case as this work:
the automated classification of commits that are \emph{security-relevant} (i.e.,
that are likely to fix a vulnerability). They train two independent classifiers
on the patch introduced by a commit (\emph{Patch Classifier}) and the log
messages (\emph{Message Classifier}), without relying on information from
vulnerability advisories. Inspired by the \emph{naturalness}
hypothesis~\cite{surveyAllamanis,hindle2012naturalness}, both classifiers treat
their input as documents written in natural language (\emph{code-as-text}), and
classify them using established natural language processing (NLP) methods. Their
results are combined with a voting mechanism flagging a commit as
security-relevant if either model does.
To process code changes in a commit, they keep only added and deleted lines of the commit and then treat these as natural language. In our experiments, as a baseline, we use token based models that follows a similar approach.

Unlike the work in~\cite{sabetta2018practical}, we focus exclusively on
the code as a source of information (that is, we do not use the commit message nor text in natural language, coming, e.g., from code comments). This makes
our method applicable to the detection of \textit{silent fixes} (security fixed that are committed to source code repositories without a specific mention of their security-relevance in the commit message and for which there is no explicit announcement through an advisory).

To the best of our knowledge, the state of the art in academic research for
source code distributed representations has mostly focused on the representation
of \emph{methods} rather than whole \emph{commits}~\cite{code2seq,code2vec}. There seems to still be a
lack of a more global representation for code, as code commits can be composed
of changes in multiple methods located in multiple files.

%As previously described,  uses a bag of aggregated contexts from the
%AST of a code snippet. The work in~ generates a sequence of
%distributed representations to represent a method rather than a single vector.

More importantly, the scope of these two works~\cite{code2seq,code2vec} is restricted to the representation
of a code snapshot, and does not consider the temporal component
necessary to represent changes in code commits (differences between two code snapshots).

The work in~\cite{representedits} does include a temporal component, as it aims
to represent edits in source code, however its scope is restricted to small edits of only a few lines of code, which would not meet the requirement of our
application scenario (Section~\ref{sec:motivation}) for which a representation of entire commits is necessary.
%  be insufficient to represent the vast
% majority of code commits.

\begin{figure*}[t]
	\centering
	\includegraphics[width=\linewidth]{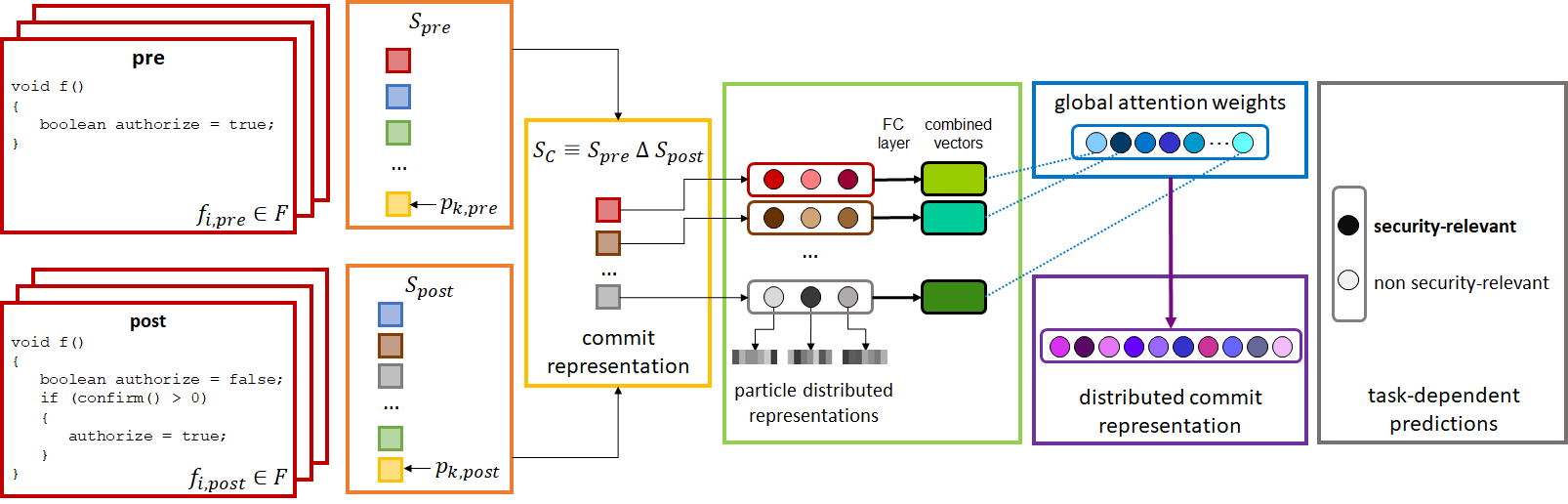}
	\caption{\ptov: the prior ($f_{pre}$) and posterior ($f_{post}$) versions of all code-relevant files in a commit, $C$, are transformed into contexts through an AST-based code representation, generating both $S_{pre}$ and $S_{post}$.
		The commit representation, $S_C$, is computed as the symmetric difference between $S_{pre}$ and $S_{post}$ and is provided as the input to a neural network.
		In this diagram, the exemplified task is that of classification of security and non-security relevant commits.}
	\label{fig:p2v_diagram}
\end{figure*}

\section{\ptov}

Finding the best-suited representation of code in machine learning frameworks is
an open research question. The survey in~\cite{surveyAllamanis} classifies the
representation of code into three main categories:
\begin{itemize}
	\item[-] \textit{token-level models} treat code as a sequence of tokens
	in a similar way as traditional natural language processing (NLP) techniques
	represent text as a sequence of words in a given language
	\item[-] \textit{syntactic models} leverage on the
	underlying structural information of code through their abstract syntax tree
	(AST) representations
	\item[-] \textit{semantic models} represent code as a graph
	generalizing both token-level and syntactic models
\end{itemize}

The concept of \textit{word embeddings}, made popular by the work in ~\cite{word2vec}, allowed a breakthrough
in many NLP-related tasks. Over the last few years, approaches inspired on the same concept are emerging in the domain of source code analysis. The work in
\cite{surveyCodeEmbeddings} presents a survey of different works that use the
concept of embeddings at different granularities of code.

In this work, we introduce a method to represent source code \emph{changes} (such as those contained in the commits of a source code repository).
Differently from the approaches that represent a static snaphot of code~\cite{code2vec}, or that represent changes in small code fragments~\cite{representedits},
we focus on representing full commits, which can contain changes across multiples methods, classes, and even files.
%  and which implies a dynamic version due to its temporal
% component.
% We further formalize and generalize this concept
% by introducing a commit representation which is representation agnostic.

% \as{not sure how to position the following; shall we drop it?}
% We exemplify the use of the our \textit{code commit} representation
% by applying it for the identification of security-relevant commits.
% The work in \cite{sabetta2018practical} had introduced the concept of a code \textit{patch}
% to describe the lines of code modified by a commit. Their work relied on a \textit{token-level model} of code.

\subsection{Commit Representation}

Our method uses \codetovec as a basic building block: for a given commit, we
extract all the methods that are changed and use the same preprocessing steps as \codetovec to extract a set of paths over the AST (\emph{context} in the terminology of~\cite{code2vec}); we then discard the contexts that are identical in the code before and after the commit, and use the remaining paths as the basis for the commit representation\footnote{Because our focus is on investigating the representation of source code, we do not consider commits that change only non-code files (e.g., metadata, readme, changelogs, documentation, etc.) or that change only comments.}.

% More formally, we define a code-relevant file, $f \in F$, as one with an extension of the
% programming language of a given project (e.g. \textit{.java} for Java,
% \textit{.py} for Python or \textit{.rb} for Ruby). Under this definition, text
% files with \textit{.txt} extension and other non-code relevant files are not
% included in $F$.

More precisely, let a code commit, $C$, be defined as a change in the source code of a given
project in a set of files $f_i \in F$, where $i \in [1..I]$,  where $I$ is the number of files changed within \textit{C}.
% Non-code changes such
% as inclusion, removal or modification of comments in $f_i$ are not considered.

The concept of a commit implies an prior and a
posterior version of files $f_i$, which we denote as
$f_{i, pre}$ and $f_{i, post}$ respectively.

Analogously to textual tokens in token-based representations, our model uses
paths constructed traversing the abstract syntax tree (AST) of each method
changed in $C$. Consistenly with the terminology of~\cite{code2vec},
we call \emph{contexts} the triplets of two terminal nodes and their connecting path on the AST.

% For our experiments, we have to define in both token-based representation and AST representation the smallest element, that we name a particle, in a given form of code representation for a code snippet or method, $m_j \in M$ for $j \in [1..J]$ with $J$ being the number of methods changed in a commit. In the case of token-based representation, we define this element as the modification of a line in the code changes, similarly to \cite{sabetta2018practical}.

%We define a \textit{particle} as the smallest element in a given form of code
%representation for a code snippet or method, $m_j \in M$ for $j \in [1..J]$
%with $J$ being the number of methods changed in a commit. In this work, a \textit{particle} can represent a
%token in a token-based representation of code, or a \textit{context} as defined by~\cite{code2vec}, when using an AST representation of code.

Let the union of all the contexts of the prior versions of all
methods $m_{1..J, pre}$ in all files $f_{1..I, pre}$ in commit $C$ be defined as
$S_{pre} = \{ p_1, p_2, ..., p_k \}$ and the union of all the contexts of the
posterior versions of all methods $m_{1..J, post}$ in all files $f_{1..I,
	post}$ in commit $C$ be defined as $S_{post} = \{ p_1, p_2, ..., p_k \} $. We then define the set of contexts describing commit $C$ as the symmetric
difference between $S_{pre}$ and $S_{post}$:

\begin{equation}
S_{C} = S_{pre} \Delta S_{post} \equiv \{p: p \in S_{pre} \cup S_{post}, p \notin  S_{pre} \cap S_{post} \}
\end{equation}

Intuitively, the symmetric difference $S_{C}$  between the two sets of contexts contains the contexts that have been changed in the commit $C$.

$S_C$ is the input provided to the neural network architecture that yields a distributed representation of the code changes performed in commit $C$. In order to generate meaningful representations, the neural network typically
requires large amounts of data to be trained on. Unfortunately, in many applications the data available
is not sufficient. In these cases,
\textit{transfer learning} techniques are applied, where the network is pre-trained on a similar task for which large amounts of data are available, often called the \textit{pretext task}, and then fine-tuned on the \emph{target
task} using a smaller dataset.

%\as{This breaks the flow a bit...}

In the following sections, we evaluate the performance of \textit{token-based} and
\textit{AST-based} commit representations for the classification of security-relevant commits. In the follwing,

\begin{itemize}
	\item[-] \textbf{\textit{Token-based commit representation}} will refer to the use of addition of removal of lines
	 in code as \textit{particles}.
	Nevertheless, the raw tokens inside these lines were preprocessed by splitting composite tokens (e.g. camel-case identifiers)
	and by removing stopwords, non-alphanumeric characters, integers and single letter values, as well as reserved
	keywords from the Java language.
	\item[-] \textbf{\textit{AST-based commit representation}} will refer to the use of \textit{contexts} as \textit{particles}. We consider the definition of
	~\cite{code2vec} of \textit{context}: a triplet of two terminal nodes and their connecting path from the abstract syntax tree (AST)
	of a code snippet.
\end{itemize}

\subsection{Research questions}
%\rcl{EXPECTED CONTENTS: Research questions, tokens vs ast vs pretraining}

We define a performance baseline by using token-based commit representations. We first replicate the work of \cite{sabetta2018practical} by
treating code as a collection of tokens in a \textit{bag of words (BoW)} approach coupled with a
support vector machine (SVM) classifier. To obtain a stronger baseline, we improve upon that model by adopting a method capable of interpreting code as a \emph{sequence} of tokens (and not just as a set, asn in BoW); we use a long short-term memory (LSTM) network for this purpose.

Following the creation of a baseline with token-based commit representations, we explore the following research questions:

\begin{itemize}

	\item[$\bigstar$] \textit{\textbf{RQ1} - Is there an added value in using syntactic models as opposed to token-level models to construct code representations?} To answer this question, we compare the performances of the models trained on the token-based representations in the baseline to an AST-based representation.
	 % As previously described, syntactic models Can an AST-based representation outperform a token-based code representations We tackle this research question by comparing the  Without external data
	\item[$\bigstar$] \textit{\textbf{RQ2} - Does pre-training a neural network on a very big ($>10^6$ samples), but loosely related, dataset provide an increase in performance for the target task?} Training on an extensive database, sometimes in the order of millions of samples, is the basis for \textit{transfer learning} techniques which has been widely applied in other domains. We use the network from \cite{code2vec} trained on a dataset of 12M samples to predict method names for our task of predicting security-relevant commits.
	\item[$\bigstar$] \textit{\textbf{RQ3} - Does pre-training a neural network on a smaller dataset ($>10^4$ samples rather than $10^6$), but on a highly relevant task provide an increase in performance for the target task?} We compare the results obtained while answering RQ2 to those obtained by pre-training a network on a pretext task which we hypothesize is more closely related to our target task. We use a large, but not massive dataset,(tens of thousand data points, as opposed to millions), to train a classifier on this pretext task.
\end{itemize}

\section{Datasets}
\label{sec:dataset}

In order to run experiments that could answer the research questions outlined in
the previous section, we took an existing dataset of security-relevant commits
as a starting point, and we complemented it with additional commits that we
mined automatically as explained below. Also, to tackle the \textit{pretext}
task, we constructed a dataset of Jira tickets obtained mining hundreds of projects from the Apache Software Foundation (ASF).

%\as{It is worth noting that, since this work focuses on representing \emph{code changes} the commits that modify only non-code file (e.g., metadata files, ``readme'' files, documentation, changelogs, etc.).}

% The following sections describe the details of these datasets:
%Furthermore, changes involving addition or removal of full methods in Java files remain also out of the scope of this work.

\subsection{Security-relevant commit dataset}
\label{sec:securitydataset}
The security-relevant commit dataset used is a combination of two sources:

\begin{itemize}
	\item[-] \textbf{\textit{Manually curated dataset.}} We use the dataset introduced
	in~\cite{ponta2019dataset} where each entry represents a commit that contributes
	to fixing a vulnerability (so-called \emph{fix-commits}). The dataset covers a representative
	sample of OSS projects of practical industrial relevance.
	%: these projects were identified
	%based on an analysis of the data collected at \textbf{[COMPANY NAME]} while operating an
	%internal open-source vulnerability assessment tool for a period of about four years,
	%during which the tool was used for hundreds of thousands of security scans. Most of
	%the open-source projects included in the dataset are hosted on \textsf{\small GitHub.com}
	%(or a mirror of their official repository is available there).
	%The dataset maps 624 publicly disclosed vulnerabilities affecting 205 distinct open-source Java projects used in \textbf{[COMPANY NAME]} software (either products or internal tools) onto the 1282 commits that fix them.
	This dataset was constructed and manually curated, monitoring the disclosure of security advisories,
	not only from the NVD, but also from numerous project-specific Web pages.
	%	\rcl{VALID COMMITS: POS 573 NEG 572 TOTAL 1145}

	\item[-] \textbf{\textit{Mining commits from Jira issues.}} We retrieved, from the Jira system of each of the OSS projects in the
	manually curated dataset,
	%\as{For Antonino: correct if wrong, did we mine the tickets of all Apache????}
	all the issues that were manually tagged as ``security-related'' (that is, that had labels such as: \textit{security}, \textit{authorization},
	\textit{authentication}) by the user or developper who created the issue.
	We then  extracted the commits linked to these issues (via a reference to the issue identifier found in the commit message).
	%	\rcl{VALID COMMITS: POS 402 NEG 403 TOTAL 805}

	%The second source is using the Jira issues dataset that we built. Each issue can have manually created
	%labels attached when created. This label gives us precise information about the nature of the issue.

	% We managed to extract 1010 issues containing one of these labels: 747 for security, 223
	%for authentication and 40 for authorization. From these issues, we extracted 402 positive commit examples
	%that can be processed by the code2vec model (which have java files and valid method changes). We randomly
	%sampled 403 commit linked to jira issues and labeled them as negatives.
\end{itemize}

Both of the previously described sources consist of only positive samples for
our problem formulation so we augment the dataset with an equal number of
negative instances (non-security commits). For each positive instance $p$ from
repository $R$, we take a random commit from $R$ and, under the assumption that
security-relevant commits are rare compared to other types of commits, we
treated these as \textit{negative} examples. To avoid including obvious outliers
(extremely large, empty, or otherwise invalid commits), a manual review of these
commits was performed, supported by ad-hoc scripts and pattern matching
(similarly to~\cite{zhou2017automated}). These patterns are used to speedup the
manual review by searching for patterns in the commit messages that would
indicate with high probability that the commit was security-relevant.

Only the commits with code changes are kept (i.e., changes in non-code files, or comments within the code are not considered). This results in a dataset
containing a total of 1950 commits, with 975 positive instances (security related) and 975 negative instances (not security related).

\subsection{Jira Ticket Dataset}

Jira is an issue tracking system developed by Atlassian which allows developers and users of a given software to manage issues, bug reports, and development tasks in general.
To create our labeled dataset of commits, we considered all the Apache Software Foundation projects written in Java (195 out of 264) and collected the Jira issues associated to them.
A large percentage of Apache projects use the Jira platform for their issues.
%  Out of 264 Java projects listed on the ASF website, we considered the 195 projects whose main programming language is Java (almost all of which use the Jira issue tracking sistem).
%We extracted Jira tickets from the 195 Apache Java projects which are listed on Table \ref{table:project_list}.
% A Jira ticket contains the following elements:

%There are other bug tracking softwares, such as Bugzilla, which are also open source.
%In this work we focus on Jira and not Bugzilla because of it directly provides a representation of the severity of an issue,
%by giving the priority of the issue. One of the other reasons is that it was easier to scrape the Jira website than the Bugzilla.

% \begin{itemize}
% 	\item[-] \textbf{Identifier}: composed of an abbreviation of the
% 	corresponding Apache project and a number. Each identifier corresponds to an
% 	unique Jira issue.
% 	\item[-] \textbf{Type}: There exist 20 ticket types e.g. bug, question, improvement
% 	\item[-] \textbf{Priority}: For almost all projects, the existing 5 types of
% 	priorities in increasing order of importance are: Trivial, Minor, Major,
% 	Critical and Blocker.
% 	\item[-] \textbf{Description}: A written explanation of the ticket in plain
% 	natural lanugage.
% 	\item[-] \textbf{Discussion}: A discussion of users about this issue.
% \end{itemize}

% The list is non exhaustive.
The projects hosted in the Apache Software Foundation (ASF) where chosen because of their industrial relevance; also, the majority of the ASF projects follow the practice of linking source code commits to Jira tickets by including
the identifier of one or more tickets in the commit message.
By mining source code repositories,
we could construct a dataset of commits mapped onto the corresponding
tickets. A commit linked to a ticket is considered as the fix of the issue
described in the ticket. For this work, the pretext task we consider is the prediction of the priority of tickets based on the code changes in the corresponding commit(s). Other pretext tasks could be defined, based on other attributes of the Jira ticket that can serve as labels. Using
ticket priority, allows us to include commits that are not linked to a
Jira issue (we assign these commits to the \textit{NoTicket} class).

\begin{figure}
		\centering
		\includegraphics[width=\linewidth]{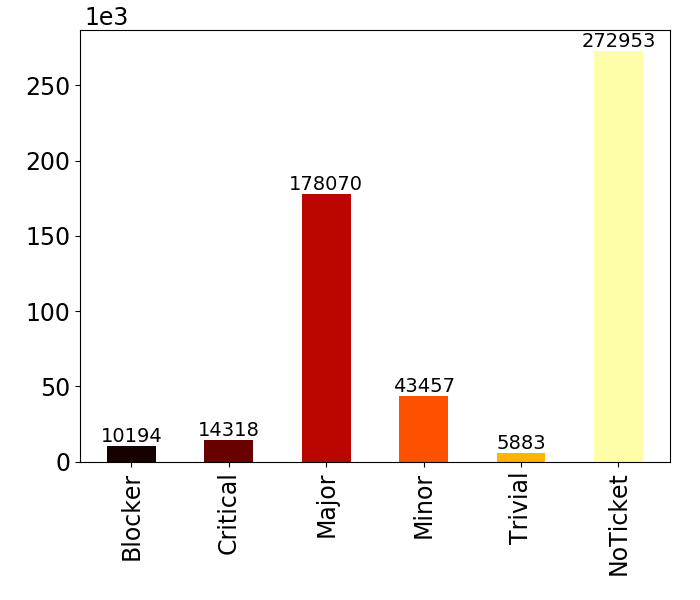}
		\captionof{figure}{Jira Ticket Priority dataset distribution.}
		\label{fig:Jira_ticketPriority}
\end{figure}

In our experiments, we keep only consider projects adhering to the five priority class convention: Trivial, Minor, Major,
Critical and Blocker.
Table \ref{table:nb_commit} shows the number of commits and their respective priority classes that we were able to
collect. The distribution of priority labels for our dataset is shown in Figure \ref{fig:Jira_ticketPriority}.

\begin{table}
	\footnotesize
	\begin{tabular}{lrrr}
		\hline
		&\textbf{References found} & \textbf{Valid commits} & \textbf{Proportion} \\
		\hline
		Blocker & 30,001  & 10,194& 0.34\\
		Critical  & 39,868 & 14,318& 0.36\\
		Major & 463,556 & 178,070& 0.38\\
		Minor & 123,252 & 43,457& 0.35\\
		Trivial & 21,222 & 5,883& 0.28\\
		NoTicket & 1,049,636 & 272,953 & 0.26\\
		\textbf{Total} & \textbf{1,727,535} & \textbf{524,875}& \textbf{0.30}\\

		\hline
	\end{tabular}
	\normalsize
	\captionof{table}{Number of commits found for each priority and remaining number of commits in the final dataset}
	\label{table:nb_commit}
\end{table}

\label{sec:dataset}

\begin{figure*}[t]
	\noindent\begin{minipage}[t]{\textwidth}
		\centering
		\begin{tabular}{llllll}
		\hline
			                              &        Precision (\%) &           Recall (\%)&               F1-score &         Accuracy (\%) & PR-AUC ($10^{-2}$) \\
		\hline
			                           BoW-SVM &  66.09 $\pm$ 2.41 &  73.13 $\pm$ 1.14 &  69.41 $\pm$ 1.64 &  67.74 $\pm$ 2.19 & 66.36$\pm$1.51 \\
			                              LSTM &  70.91 $\pm$ 4.15 &  68.71 $\pm$ 3.08 &  69.63 $\pm$ 1.42 &  70.00 $\pm$ 2.24 & 74.70$\pm$2.69 \\
			       \ptov (no pre-training) &  69.41 $\pm$ 1.70 &  77.22 $\pm$ 3.19 &  73.08 $\pm$ 2.11 &  71.59 $\pm$ 1.98 & 82.51$\pm$1.57 \\
			  \ptov (code2vec pre-trained) &  70.04 $\pm$ 1.52 &  76.09 $\pm$ 3.30 &  72.90 $\pm$ 1.73 &  71.75 $\pm$ 1.54 & 82.11$\pm$1.93 \\
			      \ptov (Jira pre-trained) &  72.01 $\pm$ 2.69 &  78.16 $\pm$ 1.16 &  74.92 $\pm$ 1.26 &  73.80 $\pm$ 1.92 & 84.41$\pm$1.42 \\
		\hline
		\end{tabular}

		\captionof{table}{Classification performance ($\mu\pm\sigma$) of models on the \textit{security-relevant commit dataset} (5-fold cross-validation).}
		\label{table:res_kfold}
		\includegraphics[width=0.8\textwidth]{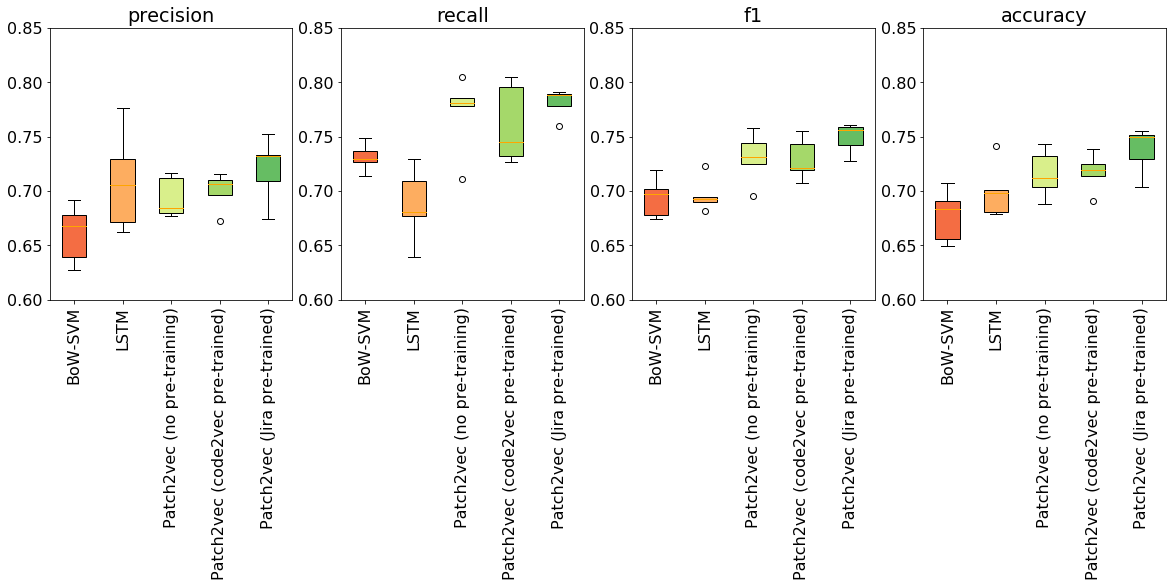}
		\captionof{figure}{Boxplot visualizations of performance metrics of models on 5-fold cross-validation on the \textit{security-relevant commit dataset}.}
		\label{fig:kfold}
	\end{minipage}%
\end{figure*}

\section{Experiments}
%\rcl{EXPECTED CONTENTS: Self contained experiments: method/result}

The following section describes the use of both token-based and AST-based commit representations, $S_C$, for the
classification of security-relevant commits. Table~\ref{table:res_kfold}  and Figure~\ref{fig:kfold} summarize the precision, recall, F1-score, accuracy and area under the precision-recall curve (PR-AUC) metrics obtained after performing 5-fold cross validation on the security-relevant commit dataset.

%\begin{table*}
%	\centering
%	%	\resizebox{0.5\textwidth}{!}{%
%	\begin{tabular}{lllll}
%	\hline
%		&      Precision &         Recall &             F1-score &       Accuracy \\
%	\hline
%		%			          BoW-SVM (with comments) &  65.44\% $\pm$ 2.05 &  73.33\% $\pm$ 0.98 &  69.13\% $\pm$ 0.99 &  67.23\% $\pm$ 1.67 \\
%		BoW-SVM &  66.09 $\pm$ 2.41\% &  73.13 $\pm$ 1.14\% &  69.41 $\pm$ 1.64\% &  67.74 $\pm$ 2.19\% \\
%		%	      LSTM(with comments) &  69.16\% $\pm$ 2.56 &  71.49\% $\pm$ 3.45 &  70.24\% $\pm$ 2.16 &  69.74\% $\pm$ 2.10 \\
%		LSTM &  70.57 $\pm$ 3.10\% &  67.39 $\pm$ 4.54\% &  68.82 $\pm$ 2.59\% &  69.54 $\pm$ 2.27\% \\
%		\ptov architecture &  68.38 $\pm$ 2.11\% &  77.13 $\pm$ 2.79\% &  72.45 $\pm$ 1.69\% &  70.67 $\pm$ 1.87\% \\
%		\ptov pre-trained &  69.84 $\pm$ 1.13\% &  76.61 $\pm$ 2.79\% &  73.04 $\pm$ 1.33\% &  71.75 $\pm$ 1.10\% \\
%		Jira pre-trained &  \textbf{70.71 $\pm$ 1.48\%} &  \textbf{79.69 $\pm$ 2.11\%} &  \textbf{74.92 $\pm$ 1.47\%} &  \textbf{73.33 $\pm$ 1.52\%} \\
%	\hline
%	\end{tabular}
%	%	}
%	\caption{\label{table:res_kfold} Classification performance of models on 5-fold cross-validation on the \textit{security-relevant commit dataset}.}
%\end{table*}

%\begin{figure*}[ht]
%
%	\centering
%	\includegraphics[width=0.8\textwidth]{box_plot_kfold.png}
%	\caption{Box plot of 5-fold experiment \rcl{update model names, rescale all axes to 0.5-1.0}}
%	\label{fig:kfold}
%
%\end{figure*}

\subsection{Baseline} %: token-based commit representation
We compare two token-based commit representations for the classification of security-relevant commits
which will serve as a baseline to compare the rest of the proposed approaches.
For both approaches, a dictionary size of 20,000 tokens was used.

%We firstly present two token based approaches, one being a support vector machine model combine
%with a bag of words and the other being an lstm.
%For these models, we pre-process the data the same way.
%For each commit in the dataset, we have extracted the method declarations and the classes constructor
%for each of the java file in the pre and the post version of the commit.
%At this point, we tried two
%configurations: one where we keep comments in method declarations and another on where we discard them.
%To keep the relevant information of the commit, we keep from methods only the lines that have been added
%or removed in the commit.
%The resulting example is the concatenation of the lines of all methods in the commit.

\subsubsection{Bag of Words and SVM}
Similarly to the approach presented in~\cite{sabetta2018practical}, the token-based commit representation, $S_C$, is
vectorized as a bag of words (BoW). The vectors are then fed into a support vector machine (SVM) with a linear kernel.
For ease of replication, we show pseudocode for this implementation in Listing \ref{BoW+SVM model} using the Sci-kit learn
framework. Variable $d$ refers to the dictionary size, and $df$ is a dataframe containing in column \textit{Sc} the
lines of code remaining after performing the symmetric difference between the two file versions.

As shown in Table~\ref{table:res_kfold}, this approach yields an average F1-score of $69.41 \pm 1.64\%$ and an average PR-AUC of $66.36 \pm 1.51$ across 5 folds.
%We firstly define an easy machine learning model based on bag of words (BoW) and support vector machines
%(SVM). Each commit example of the dataset is preprocessed before being fed to the model.
%To train the model, we vectorize each commit example by using a bag of words with a maximum dictionary
%size of 20,000.

\noindent
\begin{minipage}{\linewidth}
\begin{lstlisting}[language=Python, caption={BoW+SVM model}, label={BoW+SVM model}]
v = CountVectorizer(analyzer = "word", max_features = d)
v.fit(list(df['Sc'].values))
X_train = vectorizer.transform(df_train['Sc'].values)
clf = SVC(kernel='linear', probability=True)
clf.fit(X_train, Y_train)
\end{lstlisting}
\end{minipage}

\subsubsection{LSTM}
Here we represent $S_C$ as a sequence of tokens of maximum length of 500 tokens. Each element in the sequence will
be embedded to a vector size of 32. This sequence will be fed into a long-short term memory (LSTM) network.
%Each token in an example is transformed to an identifier to form a sequence. The
%sequence is then padded to a maximum length of 500. Each identifier is associated to a vector of size 32. The
%vectors sequence is given in input to the LSTM model.
The model is composed of an LSTM layer of 100 units and a dense layer with a sigmoid activation function.
The loss function used is binary cross entropy, and the Adam
optimizer.
%We train each LSTM model on 5 epochs.
The code in Listing \ref{lst:LSTM} demonstrates the
implementation using the Keras framework. Variable $d$ refers to the dictionary size, $e$ to the embedding size, $h$ to the number of units
in the LSTM layer and $s$ to the maximum sequence length.

This approach yields an average F1-score of $69.63 \pm 1.42\%$ an an average PR-AUC of $74.70 \pm 2.69$ across 5 folds.

\noindent
\begin{minipage}{\linewidth}
\begin{lstlisting}[language=Python, caption={LSTM model}, label={lst:LSTM}]
model = Sequential()
model.add(Embedding(d, e, input_length=s))
model.add(LSTM(h, return_sequences=True))
model.add(Flatten())
model.add(Dense(1, activation='sigmoid'))
model.compile(loss='binary_crossentropy', optimizer='adam')
\end{lstlisting}
\end{minipage}

 %%\vspace{0.2cm}

In the experiments performed, treating $S_C$ as a sequence of tokens yielded a
non conclusive increase of $0.22$ percentual points in F1-score with respect to
a bag of words approach. Nevertheless, when comparing the PR-AUC scores, we
observe an increase of $8.34$ percentual points. It should be reminded that the
precision and recall values reported are those computed for a classification
threshold of 0.5, whereas a precision-recall curve shows the tradeoff between
precision and recall for different thresholds. Therefore, a PR-AUC will provide
a more comprehensive performance measurement in the case where 0.5 would not
belong to the optimal classification threshold for the given task.

Given these results, we can state that treating code as a sequence of tokens,
which provides more information than the mere collection of tokens, increases
the performance for the prediction of security-relevant commits. In the next
sections, we compare this baseline to AST-based commit representations. These
representations contain information about the structure of code and better
capture the dependencies between related tokens, despite them being far apart in
the code. We hypothesize that AST-based representations will outperform
token-based representations for our target classification task.

%The increase in
%		 precision was of $4.53$ percentual points but was countered by a decrease in recall of $4.32$ percentual points. While this might initially seem counterintuitive, a possible explanation could be the nature of our commit representations. A commit is represented solely by the changes in code that occurred, therefore, unless the commit was composed of long, contiguous changes in the code, the commit representation might not be an
%		 \rcl{ interpretable sequence of code. If the commit was composed of short, spaced out changes then seeking to interpret the changes as a sequence would be meaningless, almost equivalent to a mere collection of tokens.  This could explain the very similar performances of both algorithms. AUC PR curve, lstm comes ahead?? }

\begin{figure}[htp]
	\centering
	\includegraphics[width=\linewidth]{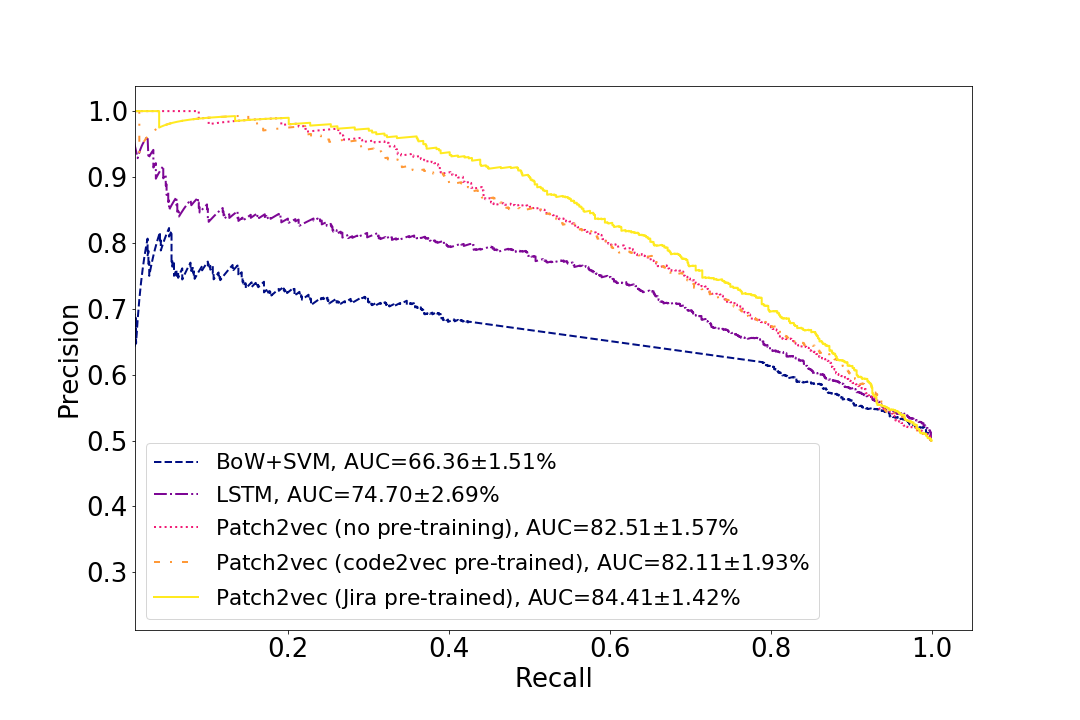}\hfill
	\caption{Precision-recall curves for each model averaged over 5-fold cross validation results. }
	\label{fig:pr_kfold}
\end{figure}

\subsection{AST-based commit representation}
\label{sec:AST-experiments}
We compare the use of an AST-based commmit representation for the classification
of security-relevant commits with and without the use of external databases to
pre-train the network. In all the following cases, we restrict the size of
$S_C$ to maximum 500 contexts. Commits with less contexts are padded with
empty values and for those with higher number of contexts, an sample of 500 contexts
is randomly chosen. We empirically found this number to be suitable to
represent a code commit which can be composed of changes in multiple methods in
multiple files.

%An AST-based commit representation is obtained by
%will refer to the use of \textit{contexts} as \textit{particles}. We consider the definition of
%~\cite{code2vec} of \textit{context}: a triplet of two terminal nodes and their connecting path from the abstract syntax tree (AST)
%of a code snippet.

%\rcl{We input the commit representation, $S_C$ ... all of the next models will have an ast-based commit representation}

\subsubsection{\textbf{\ptov (no pre-training)}}

We use the \textit{security-relevant commit dataset} described in Section
 \ref{sec:securitydataset} to train a classifier. We trained a model with a
 similar architecture as the one described in~\cite{code2vec}. The last layer of
 the original model architecture is replaced by
%a fully connected layer of 64 neurons (with batch normalization and ReLU activation) and
a binary softmax layer as output corresponding to the two classes, security-relevant and non security-relevant commit.
All network weights, including embeddings of terminals and paths, are initialized randomly. The model was trained using binary cross-entropy as
a loss function and the Adam optimizer. The average F1-score obtained by this approach is of $73.08 \pm 2.11 \%$ and the average PR-AUC score is of $82.51 \pm 1.57 \%$ over 5 folds .

%%\vspace{0.2cm}
\begin{itemize}
	\item[$\bigstar$] \textit{\textbf{Answer to RQ1}} From the results in Table \ref{table:res_kfold}, it can be seen that training a model on
	AST-based commit representations came with an increase of at least $3.45$\% in F1-score and $7.81$\% in PR-AUC score with respect to their
	token-base counterparts.
%	 While the precision score was very similar to the token-based approaches, we can see an increase of
%	at least $4.09$ percentual points in the recall.
	Furthermore, by inspecting Figure \ref{fig:pr_kfold} it is clear that training
	on an AST-based representation translates to substantially higher precision performances for any given recall value. We
	believe this increase in performance is due to the structural information of code contained in an AST-based representation.

\end{itemize}

%\rcl{HERE OR IN DISCUSSION? talk about comments/no comments? maybe even in conclusion as further work, a token based algorithm for code comments and an ast model for code itself}

\subsubsection{\textbf{\ptov (code2vec pre-trained)}}

We used the pre-trained network provided by~\cite{code2vec}. This network was trained on over 12M methods for the
prediction of method names and the learned code-vectors have been shown to capture
semantic similarities, combinations, and analogies~\cite{code2vec}.  In a similar way as in the previous section,
the last layer of the original model was stripped and replaced by a binary softmax layer. All the network weights were allowed to evolve
during the fine-tuning stage using the training set of the \textit{security-relevant commit dataset}. % the embedding and attention weights were allowed to evolve.
The average F1-score obtained was of $72.90 \pm 1.73\%$ and the average PR-AUC was of $82.11 \pm 1.93\%$.

% %%\vspace{0.2cm}
\begin{itemize}
	\item[$\bigstar$] \textit{\textbf{Answer to RQ2} } From the experiments performed in this work,
	we found a slight decrease of $0.18$ percentual points on the F1-score and a decrease of $0.40$ percentual points
	on the PR-AUC score averaged across five folds when pre-training
	the network on a dataset for method name prediction. This marginal decrease can be due to the very loose relation
	between the chosen pretext and the target tasks. Indeed, even though the inputs
	to the models are both composed of collection of triplets from the AST of code, there are two main differences. First,
	the original task in \cite{code2vec} aimed to predict the names of individual methods, therefore, the input to the
	network consisted of a collection of triplets which were all related to each other because they shared the same AST.
	Nevertheless, our input, $S_C$, is not composed of contexts coming all from the same AST and do not consitute a
	method on its own. The contents of our commit representation can come from multiple AST representing different
	methods located in various files of the project whose relationship between each other is not
	necessarily direct. Therefore, $S_C$ cannot be interpreted as a method. A second, and perhaps more important
	difference, is that $S_C$ does not represent a static
	version of code as would be the case of the input in the task in \cite{code2vec}.
	It represents a \textit{code commit} which inherently implies a temporal change.
	These differences can explain the results obtained.
	In the following experiment, we explore the use of a pretext task more closely related to our
	target task. Notably, we chose a pretext task where the input to the network can contain contexts
	from multiple methods but also contains the notion of a code change.
\end{itemize}

\subsubsection{\textbf{\ptov (Jira pre-trained)}}
\label{sec:classif_jira}
\begin{figure}
	\centering
	\includegraphics[width=\linewidth]{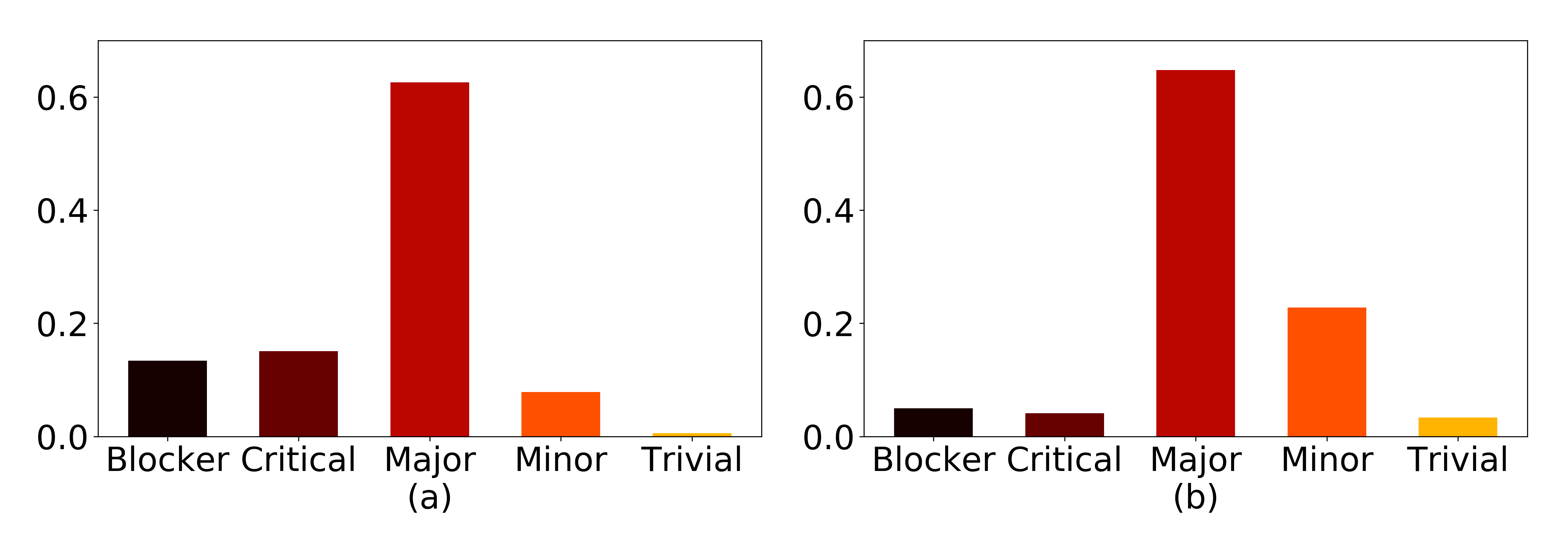}
	\caption{Proportion of ticket priorities for (a) security relevant and (b) non-security relevant commits}
	\label{fig:prio_proporions}
\end{figure}

We used the \textit{Jira Ticket Priority} dataset described in Section \ref{sec:dataset}
to pretrain our network. We hypothesize that commit fixes related to higher priority tickets are more likely to
contain fixes of important bugs, and that these could be more closely related to security-relevant fixes. This is supported by
Figure \ref{fig:prio_proporions} which shows that security-relevant commits have a higher proportion of
\textit{Critical} and \textit{Blocker} issues than non-security relevant commits. Similarly, non-security commits tend to have a higher
proportion of \textit{Minor} and \textit{Trivial} issues than positives. However, both classes contain a big majority of \textit{Major} issues.% \rcl{stat test would help}

The network used has small architecture variations from the one presented in \cite{code2vec}.
The softmax layer was replaced by a
smaller one of six classes corresponding to the six priority levels. Our dataset is split into 80\% training and 20\% test sets.\\

\textbf{\textit{Pretext task.}} Results on the classification of Jira Ticket Priority can be found on Table \ref{table:classif_pretexttask}.
Precision and recall curves are given in Figure \ref{fig:pr_curve_jira}. It can be seen that, as expected, the highest classification
performances are obtained by the predominant classes, \textit{NoTicket} and \textit{Major}. Nevertheless, the model presents micro-average
performances of 79.69, indicating an overall good performance.\\ %\rcl{i feel like the micro average would be the most relevant avg, not to
%	bias them to the more prevalent classes... actually, now im not sure XD}
	%good The model has a good micro and weighted average rate for each classification metric. Also, for classes with less examples (Blocker, Trivial, and Critical) , the model is able to have a a high precision (around 80 \%) for a reasonable recall (40\% to 60\%).

%%%\vspace{0.5cm}
\textbf{\textit{Target task.}} After pre-training on the Jira Ticket Priority dataset, the network was fine-tuned using the
\textit{security-relevant commit dataset}. Similar to the previously presented experiment, all network weights were allowed to evolve
during fine-tuning. The average F1-score obtained on the target task across 5-folds was of $74.92 \pm 1.26\%$ and the average PR-AUC score was of $84.41 \pm 1.42\%$.\\

%%\vspace{0.2cm}
\noindent\begin{minipage}[t]{\linewidth}
	\centering

	\begin{tabular}{lrrr}
	\hline
		&     F1-score &  Precision &  Recall \\
	\hline
		Blocker          &  66.47 &      73.19 &   60.88 \\
		Critical         &  69.42 &      76.13 &   63.81 \\
		Major            &  76.76 &      77.32 &   76.21 \\
		Minor            &  60.03 &      64.16 &   56.40 \\
		Trivial          &  55.52 &      67.23 &   47.28 \\
		NoTicket         &  85.76 &      83.73 &   87.89 \\
	\hline
		micro average    &  79.69 &      79.69 &   79.69 \\
		macro average    &  68.99 &      73.63 &   65.41 \\
		weighted average &  79.42 &      79.34 &   79.69 \\
	\hline
	\end{tabular}
	\captionof{table}{Model performances on the pretext task: Jira Ticket Priority classification }
	\label{table:classif_pretexttask}

	\includegraphics[width=\textwidth]{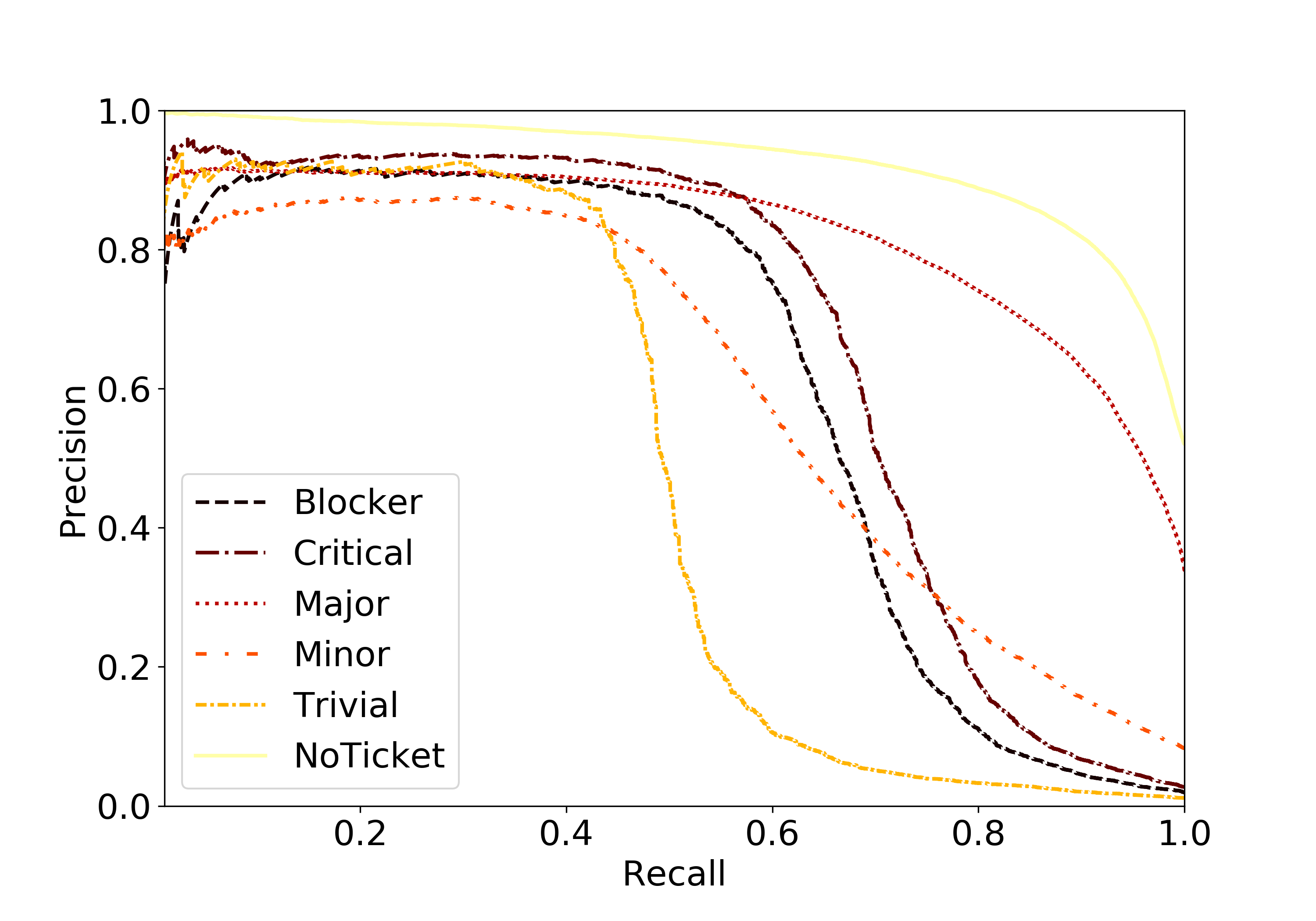}
	\captionof{figure}{Precision-recall curves for each class in the pretext task: Jira Ticket Priority classification}

	\label{fig:pr_curve_jira}
\end{minipage}%

% %%\vspace{0.5cm}

\begin{itemize}
	\item[$\bigstar$] \textit{\textbf{Answer to RQ3}} Results on Table \ref{table:res_kfold} show that pretraining on a
	smaller, yet more relevant, dataset yield an increase of $2.02$ percentual points in F1-score and $2.30$ percentual points in PR-AUC with respect to training on a loosely
	related task and of $1.84$ percentual points in F1-score and $1.9$ in PR-AUC when not using a pretext task to pretrain the network at all. These results highlight the
	importance of adequately choosing the pretext task.
\end{itemize}

\begin{figure*}
	\noindent\begin{minipage}[t]{\textwidth}
		\centering

		\begin{tabular}{lllll}
		\hline
			&      Precision &         Recall &             F1-score &       Accuracy \\
		\hline
			BoW-SVM &  63.67 $\pm$ 0.00\% &  68.27 $\pm$ 0.00\% &  65.89 $\pm$ 0.00\% &  64.66 $\pm$ 0.00\% \\
			LSTM &  68.24 $\pm$ 3.94\% &  67.74 $\pm$ 6.25\% &  67.66 $\pm$ 1.85\% &  67.66 $\pm$ 2.99\% \\
			\ptov (no pre-training) &  69.01 $\pm$ 6.37\% &  72.33 $\pm$ 8.52\% &  69.87 $\pm$ 2.03\% &  68.92 $\pm$ 3.27\% \\
			\ptov (code2vec pre-trained) &  69.89 $\pm$ 1.05\% &  72.12 $\pm$ 1.52\% &  70.97 $\pm$ 0.56\% &  70.50 $\pm$ 0.59\% \\
			\ptov (Jira pre-trained) &  72.64 $\pm$ 1.61\% &  71.10 $\pm$ 1.73\% &  71.83 $\pm$ 0.66\% &  72.12 $\pm$ 0.82\% \\
		\hline
		\end{tabular}

		\captionof{table}{Effect of random initialization in the variability of a classifier's performance (100 runs). }
		\label{table:res_100runs}

		\includegraphics[width=0.8\textwidth]{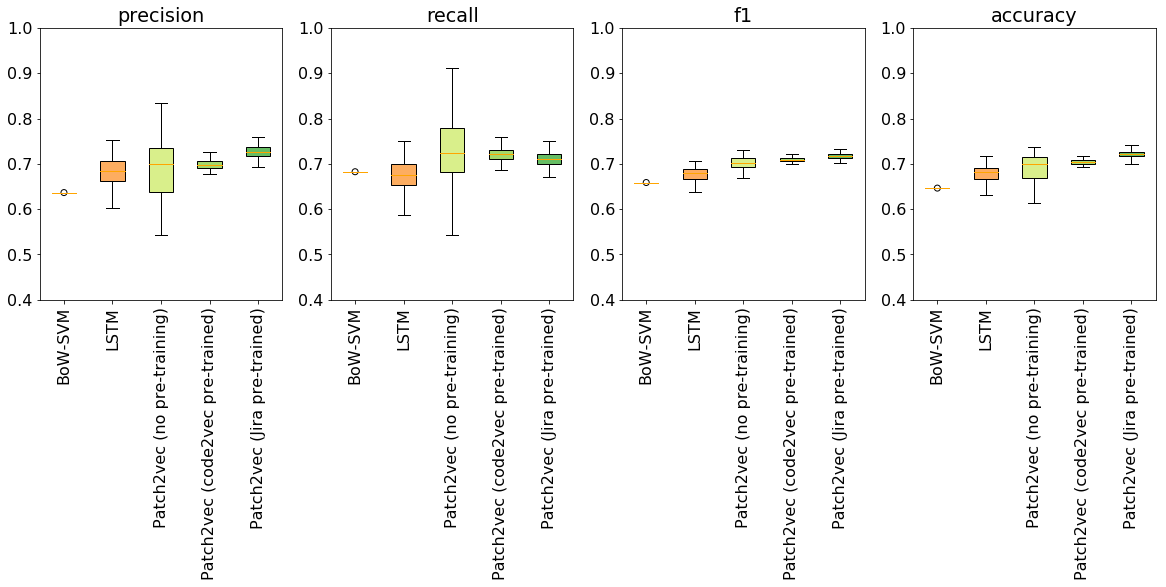}
		\captionof{figure}{Boxplots showing the high variability of performance metrics on models with high proportions of randomly initialized weights. }

		\label{fig:res_100runs}
	\end{minipage}%
\end{figure*}

\section{Discussion}

Our experiments show a superior performance for AST-based commit
representations compared to the token-based baseline, as illustrated by the larger area under the precision-recall
curves in Figure~\ref{fig:pr_kfold}. Also, in these experiments, training on a
highly relevant pretext task proved to be more effective than training on a
much larger dataset but for a losely related task.

In the rest of this section, we discuss some of the current limitations our
work and the factors that might impact the ability to generalize our findings.

\subsection{Programming language and preprocessing}

Our approach inherits some of the limitations of the \codetovec implementation,
such as the fact that only Java is supported. As a consequece, we could not use
the entire dataset of~\cite{ponta2019dataset} (which also includes
vulnerabilities affecting Python projects, and that we had to exclude).

% In order
% to tackle this, a parser for multiple languages able to generate a standardized
% AST would be needed.

Moreover, in the \codetovec implementation, the extraction of \textit{contexts} from the AST only takes into account methods (in particular, constructors or other changes outside of methods are not supported). Finally, for efficiency reasons we
did not consider methods that were completely removed or added in a commit.
Most of these limitations could be addressed with modifications to \codetovec
as well as to \ptov, which we plan to do as future work.
% It would be however simple to extend the tool to
% reduce these limitations and better exploit the available data.

\subsection{Randomness of Context Sampling}

As mentioned in Section~\label{sec:AST-experiments}, we set the maximum size of $S_C$ to 500, which means that for commits that resulted in more than 500 contexts, we randomly sampled 500 contexts.
In our experiments, this sampling was done once and kept for all runs. However, we cannot exclude that sampling different contexts for a given commit might influence the results of the classification task. We plan to address this limitation in our future work. We are considering to design a variant of our
method that could use a deterministic algorithms to reduce the number of contexts in large commits (e.g., removing very common contexts or applying expert-defined heuristics to prioritize the contexts to keep).

\subsection{Effect of Random Initialization}

Most of the models used in this work rely on a random initialization of the
weights in their networks. Different initialization values can drive the
(stochastic) optimizer to converge to different minima in the search space. This effect can impact repeatability, since different initial values could result in differences in the classifier performance. Initializing a part of the weights
in the network with pre-trained weights can reduce this effect, and we studied experimentally the variability of the classifier performance with different random initializations for each of the models.

We fixed a training ($80\%$) and testing ($20\%$) set and ran each of the models
100 times. The results are given in Table \ref{table:res_100runs} and Figure
\ref{fig:res_100runs}. Firstly, because the BoW+SVM model is
deterministic there is no variation in the performances obtained. Next,
we can see that the models with a larger proportion of randomly initialized
weights, such as the LSTM and the \patchtovec without pre-training, have a higher variability in the performance metrics. The pre-trained versions of
\patchtovec are also affected by the randomness in the initialization but because the pre-trained weights (which are a big proportion of the total number of weights) are fixed to the same value on each run, the variability in the
classifier performance is smaller.

\subsection{Pretext Task: Classification of Jira Tickets}

For this work, we defined a \textit{pretext task} to aid us solve the task of
classifying security-relevant commits through transfer learning. We chose the prediction of the priority of Jira tickets based on our experience and on the
practical observation (see Figure~\ref{fig:prio_proporions}) that security-relevant commits tend to
be associated with tickets of higher priority, while no-security-relevant
commits are associated to lower-priority tickets.

\begin{figure}[t]
	\centering
	\includegraphics[scale=0.40]{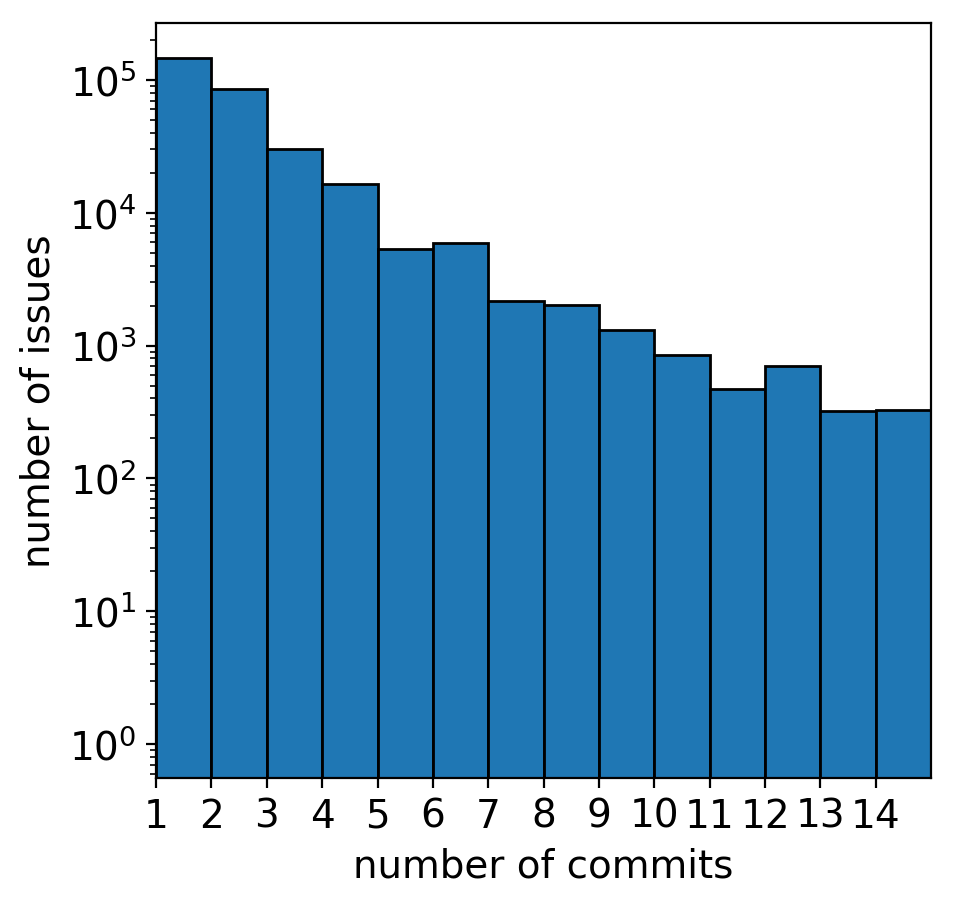}\hfill
	\caption{\label{fig:commit_per_jira}Number of commits per Jira Ticket issue.}
\end{figure}

Nevertheless, the task of predicting the priority of Jira ticket
task has multiple challenges of its own. First, there is no one-to-one
correspondence between a commit and a ticket. In fact, an issue can be
referred to in multiple commits (e.g., because a single commit may not be enough to solve the  issue). Moreover, we observed that multiple commits from different projects can refer to a same issue.
This can happen, for example, when one or multiple projects have
another project as a dependency.
If the dependency has a ticket, then a
first commit that addresses it could me made, after which other commits will be created in the dependent projects referring to the same issue. We observed this
situation in projects that belong to the same ecosystem or that are developed
by the same community (e.g., Hive, Hadoop and Spark, or Camel and ActiveMQ). Figure~\ref{fig:commit_per_jira} shows the number of commits per Jira issue.
While most of the issues are associated to only one commit, a
considerable number of issues are associated to two or more commits.
In these cases, our criterion to link tickets to the fix-commits can introduce
incorrect associations and result in noisy data, that ultimately affect negatively the training of models.

Furthermore, the ticket priority is manually set by the users, which can
have different interpretations of what a \textit{Major} or a
\textit{Minor} priority should be.
% Even in very mature project, these
% differences can arise. These are only a few reasons that make the problem of
% classifying Jira ticket priorities is very challenging.

Finally, another challenge to this task derives from the priority distributions we encountered (shown in Figure~\ref{fig:Jira_ticketPriority}). Most of
the projects have a large proportion of commits that are either associated to tickets with priority \textit{Major} or that have no associated ticket at all(\textit{NoTicket}); the other priorities concern a minority of commits.

There are also projects in which commits that reference Jira tickets are rare. For these projects, almost every issue gets labeled as \textit{NoTicket}. This could make the classification of commits from these projects trivial if the project has a very unique vocabulary. Indeed, the vocabulary of tokens, terminals and paths of the project could be very different from the global projects vocabulary. In these cases, the risk is that the model could learn the project vocabulary distribution and not its ticket priority distribution.

\begin{figure}[t]
	\centering
	\includegraphics[width=0.4\textwidth]{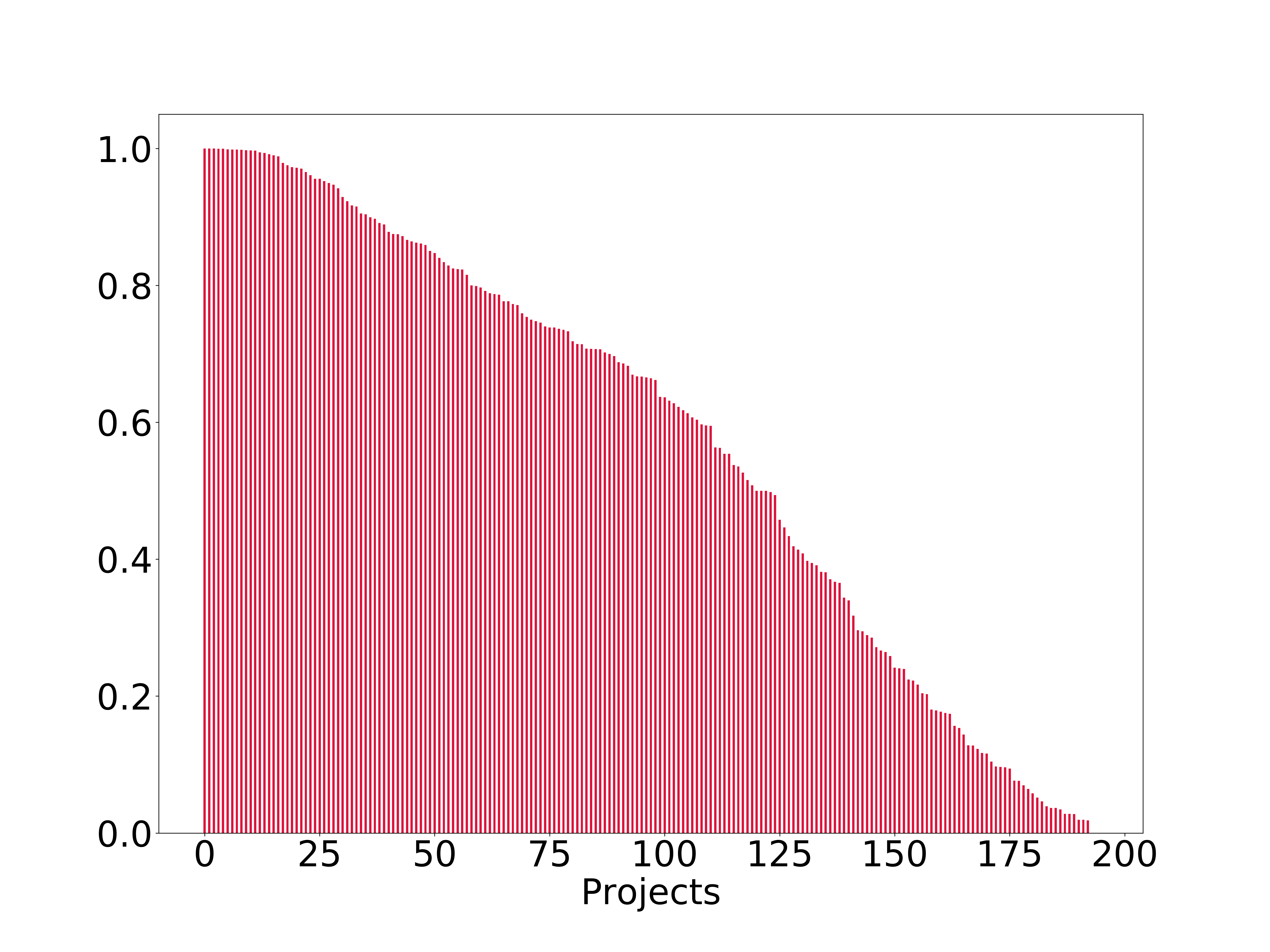}
	\caption{\label{fig:noticket_prop}Proportion of NoTicket labeled examples for each project.}
\end{figure}

%as commits from these
%projects become easily differentiable from other projects, e.g. if the terminal and path vocabularies used in project tend to be really different from the terminal and path vocabularies used in all other projects.
%In our dataset, 155 out of 195 projects have more than 10$\%$ of commits labels different from \textit{NoTicket} and  132 out of 195 projects have more than 20$\%$ of commits labels different from \textit{NoTicket}. Figure \ref{fig:noticket_prop} shows the proportion of \textit{NoTicket} labeled examples for each project. To see the impact of these projects, we tried to train \ptov on the Jira Ticket Priority classification task, but by keeping only the projects with more than a certain percentage of priority labels other than \textit{NoTicket}. We tested thresholds of 10 and 20\%. We saw a drop in classification of for the NoTicket class. There was also a minor drop for other classes. The drop could be due to less training samples. \rcl{not convinced about leaving this last section, maybe just go from different vocabularies to TSNE?, we'll see how much space we have left}

\begin{figure}[t]
	\centering
	\includegraphics[width=0.5\textwidth]{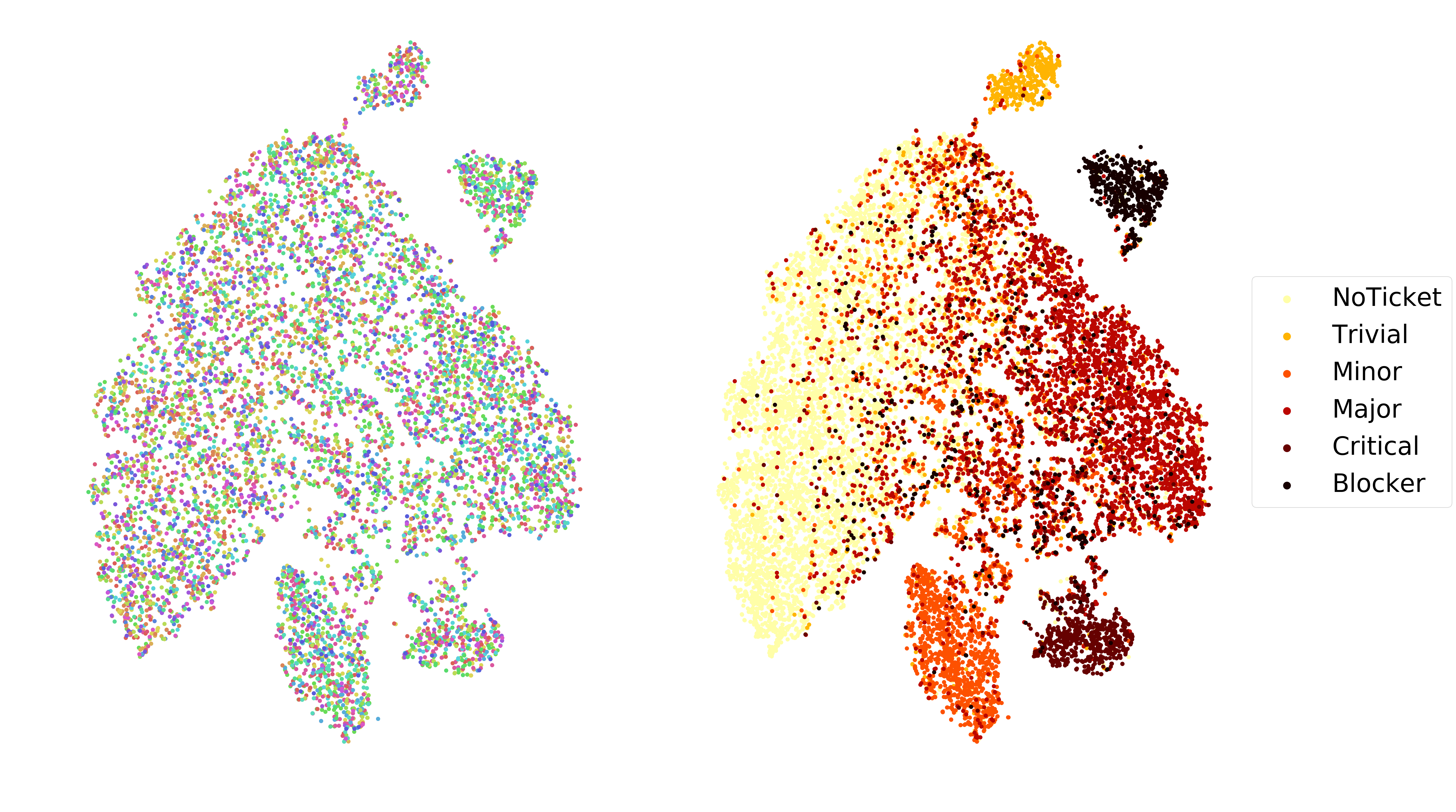}
	\caption{t-SNE plot of generated code-vectors.}
	\label{fig:tsne}
\end{figure}

To verify that this phenomenon does not happen in our study, we used t-SNE on the code-vectors of the test
dataset to generate a 2D visualization. Figure~\ref{fig:tsne} shows two plots of
a sample of code-vectors. In order not to clutter the diagram, we plot only code-vectors
from the 20 largest projects. On the right-hand plot, each color represents
a priority; on the left-hand, colors correspond to projects. We can see that
there is no clear separation or clustering of projects, meaning that the model
did not learn a \emph{distributed representation of projects}. The two biggest classes,
\textit{Major} and \textit{NoTicket}, contain many different projects and
form large, sparse overlapping clusters, showing a large intra-class distance,
but with quite a distinguishable separation between the two. The other classes form small and more dense clusters.

\section{Conclusion}
%\rcl{EXPECTED CONTENTS: restrict to certain vulnerability types}

In this work, we introduced a method to represent source code changes capable of
representing full commits,
which can be composed of changes in multiples methods and in multiple files.

We exemplify the use of the our \textit{code commit} representation by applying
it to the problem of classifying security-relevant commits. We compare the use
of token-based commit representation to AST-based commit representations. We
observed a clear superior performance for AST-based commit representations.
Furthermore, we explored the use of \textit{transfer learning} techniques by
pre-training our models on different pretext tasks. We conclude that training on
a highly relevant pretext task is to be more beneficial than training on an
extensive task with a larger dataset.

\paragraph*{Acknowledgements.} This work was partly supported by EU-funded
projects \textsc{AssureMOSS} (Grant no.~952647) and \textsc{Sparta} (Grant no.~830892).

% \bibliographystyle{acm}
% \bibliography{references}

\end{document}